\documentclass[letterpaper]{article} % DO NOT CHANGE THIS
\usepackage[preprint]{aaai2027}
\usepackage[hyphens]{url}  % DO NOT CHANGE THIS
\usepackage{graphicx} % DO NOT CHANGE THIS
\usepackage{natbib}  % DO NOT CHANGE THIS AND DO NOT ADD ANY OPTIONS TO IT
\usepackage{caption} % DO NOT CHANGE THIS AND DO NOT ADD ANY OPTIONS TO IT
\usepackage{algorithm}
\usepackage{algorithmic}

\usepackage{newfloat}
\usepackage{listings}
\DeclareCaptionStyle{ruled}{labelfont=normalfont,labelsep=colon,strut=off} % DO NOT CHANGE THIS
\floatstyle{ruled}
\newfloat{listing}{tb}{lst}{}
\floatname{listing}{Listing}

\usepackage{multirow}
\usepackage{amssymb}          % for \checkmark
\usepackage[table]{xcolor}    % for \rowcolor (gray shading)

\usepackage{booktabs}

\title{Text-Video Retrieval via Multi-Dimensional Saliency Assessment and Granularity-Aware Query Decomposition}

\author{
    Shuquan Wei\textsuperscript{\rm 1},
    Xi Chen\textsuperscript{\rm 1},
    Xu Chen\textsuperscript{\rm 1},
    Xiangyang Jia\textsuperscript{\rm 1}
}
\affiliations{
    \textsuperscript{\rm 1}School of Computer Science, Wuhan University\\
    \texttt{\{2021302111045,chenxi00233,xuchen,jxy\}@whu.edu.cn}
}

\begin{document}

\maketitle

\begin{abstract}
Text-video retrieval, which aims to bridge visual and textual modalities by learning a joint embedding space, has become a crucial task in multimodal intelligence. Despite extensive efforts to mitigate visual redundancy, previous methods typically rely on a single-aspect criterion to assess visual importance, overlooking the multifaceted spatiotemporal nature of video. In addition, encoding text into a single global embedding to align with videos compresses temporal events and spatial entities into a unified representation space, further aggravating cross-modal misalignment. To address these issues, we propose MMTI, a method that jointly mitigates visual redundancy and enables multi-grained text-video interaction to achieve accurate multi-grained semantic alignment. Specifically, a key feature selection (KFS) mechanism adaptively identifies and aggregates informative frames and patches by jointly evaluating multi-dimensional saliency and learnable importance scores, effectively compacting dense visual features and mitigating visual redundancy. Furthermore, our proposed multi-grained text-video interaction module (TVIM) employs a dynamic gating mechanism to decompose the text query into sentence, frame, and patch queries (SFP), enabling multi-grained text-video alignment. Complementary alignment at different granularities is thereby achieved. Extensive experiments on four standard benchmarks demonstrate that our method outperforms state-of-the-art methods.
\end{abstract}

\begin{figure}[t]
  \centering
  \includegraphics[width=\linewidth]{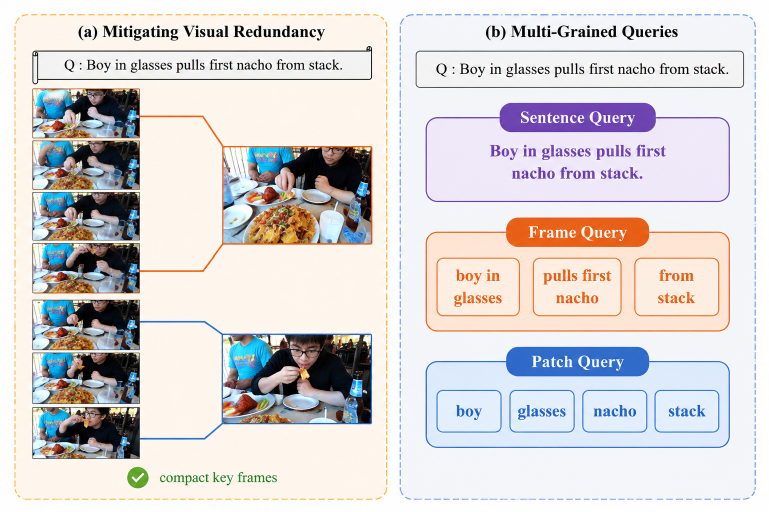}
  \caption{Illustration of our motivation. (a) Video includes highly similar adjacent frames and many repeated background areas within a frame, resulting in inherent visual redundancy. (b) Encoding the text query into a single global embedding compresses temporal events and spatial entities into a unified representation space, which significantly aggravates cross-modal misalignment.}
  \label{fig:motivation}
\end{figure}

\section{Introduction}
With the rapid growth of online video platforms, massive volumes of video data are generated daily, making effective video understanding and retrieval increasingly important~\cite{gorti2022x,ge2022bridging}. Text-video retrieval, which aims to retrieve relevant video content through natural language queries by linking visual content with semantic captions, has become a crucial task in multimodal intelligence~\cite{bain2021frozen,luo2022clip4clip}. However, unlike static image-text retrieval where a single image contains limited information, video data inherently possesses complex spatiotemporal structures characterized by high temporal redundancy and rich, dynamic semantics. Multiple objects, actions, and events unfold across frames, making it difficult to simultaneously achieve accurate cross-modal alignment and efficient retrieval~\cite{hur2025narrating}. These unique characteristics of video data require specialized techniques that go beyond simple extensions of image-text models.

Recent research has made substantial progress in text-video retrieval~\cite{le2025bima}. Existing methods can be divided into two categories. The first category of methods employs dense video embeddings to align directly with short text captions. Representative works such as CLIP4Clip~\cite{luo2022clip4clip} aggregate frame-level features into a whole video representation through mean pooling, while X-Pool~\cite{gorti2022x} aggregates features into a text-conditioned holistic video representation. Despite their effectiveness, both of the aforementioned methods process all visual features indiscriminately, inevitably suffering from visual redundancy introduced by highly similar adjacent frames and repetitive background regions. To address this, several efforts have been devoted to mitigating visual redundancy. TempMe~\cite{shen2024tempme} mitigates temporal redundancy via progressive token merging across clips, achieving superior retrieval accuracy over prior methods. V-Sparse~\cite{liu2026v} further performs text-conditioned spatiotemporal compression, retaining query-relevant visual features to suppress redundancy. However, these methods typically rely on a single-aspect criterion to assess visual importance, overlooking the multifaceted spatiotemporal nature of video. The second category of methods~\cite{hur2025narrating,le2025bima} enhances video features through text interaction. However, these methods tend to use textual features to enhance visual features and encode text into a single global embedding to align with videos of complex spatiotemporal structures. This compresses temporal events and spatial entities into a unified representation, overlooking that textual captions naturally encapsulate multi-level semantics, so that fine-grained correspondences at complementary semantic levels are not fully captured~\cite{wang2024text}.

Considering that each category of methods addresses only one facet of the problem while introducing its own limitations, an ideal solution should simultaneously mitigate visual redundancy and enable multi-grained text-video alignment. Achieving this goal requires careful consideration of two critical aspects: (1) \textbf{Select discriminative visual representations}. As shown in Figure~\ref{fig:motivation}~(a), redundant visual information must be effectively mitigated while preserving the essential semantics of the video, the key objects, actions, and events that are discriminative for retrieval. Therefore, a selection mechanism that is both adaptive and soft-aggregating is needed, rather than relying on sparse sampling or discarding. (2) \textbf{Enable multi-grained cross-modal alignment}. As shown in Figure~\ref{fig:motivation}~(b), cross-modal alignment should not be limited to global video-text matching; it should also capture complementary semantics at multiple levels, from overall content to temporal segments and fine-grained visual details. This hierarchical alignment not only captures correspondences more comprehensively, but also naturally reduces the impact of textual inaccuracies at any single granularity, as information from other levels can compensate.

Motivated by the above discussion, we propose MMTI, a method that mitigates visual redundancy and enables multi-grained text-video interaction. The core of MMTI consists of two key components. First, we design a learnable key feature selection~(KFS) mechanism that adaptively identifies and aggregates the most informative frames and patches. Specifically, KFS jointly evaluates multi-dimensional saliency and learnable importance scores, and then applies a soft assignment fusion strategy to compact the dense visual features into a concise yet informative set, thus mitigating visual redundancy. Second, we introduce a text-video interaction module~(TVIM) that decomposes the text query into sentence, frame, and patch queries~(SFP) via a dynamic gating mechanism for multi-grained alignment with the selected key visual features. The sentence query captures global video-level semantics, the frame query focuses on temporal segments most relevant to the text, and the patch query attends to fine-grained visual details. By jointly addressing visual redundancy and enabling multi-grained cross-modal alignment, MMTI achieves superior retrieval accuracy.

Our main contributions are summarized as follows:
\begin{itemize}
    \item We propose KFS that jointly evaluates multi-dimensional saliency and learnable importance scores, then applies soft assignment fusion to mitigate visual redundancy while compacting discriminative semantics.
    \item We introduce a multi-grained text-video interaction module~(TVIM) that decomposes the text query into SFP via a dynamic gating mechanism for alignment at complementary semantic levels.
    \item We design a hybrid objective function that combines bidirectional contrastive learning with multi-grained consistency regularization for improved alignment.
    \item Extensive experiments on four standard benchmarks demonstrate that our method achieves superior retrieval accuracy over state-of-the-art methods.
\end{itemize}

\section{Related Work}
\subsection{Text-Video Retrieval}
Existing text-video retrieval methods mainly aim to bridge the semantic gap between visual content and natural language captions by learning a joint embedding space. Vision-language pre-trained models, such as CLIP~\cite{radford2021learning}, are used for semantics extraction~\cite{wu2023cap4video,xu2021videoclip,zhao2022centerclip}. Building on this foundation, many methods focus on extending such models to the video domain. Representative works, including CLIP4Clip~\cite{luo2022clip4clip}, X-CLIP~\cite{ma2022x}, and X-Pool~\cite{gorti2022x}, adapt CLIP for video tasks by extracting frame-level features and employing various aggregation strategies, such as attention mechanisms to highlight text-relevant frames. Some methods introduce additional modalities, such as audio~\cite{akbari2021vatt,lin2022eclipse}, which have attracted increasing attention. These methods indicate that effective feature extraction is crucial for aligning video content with short textual queries. However, these approaches often process all visual features directly, which can introduce visual redundancy, thereby limiting retrieval performance~\cite{luo2022clip4clip,gorti2022x}. Other methods have emerged to deal with visual redundancy. V-Sparse~\cite{liu2026v} prunes redundant frames and patches via text-guided temporal-spatial semantic compression. MSIA~\cite{chen2024multilevel} suppresses redundant video frames through a frame adaptation attention mechanism. However, they often rely on a single-aspect criterion to assess visual importance, which may not fully capture the multifaceted spatiotemporal nature of video content.

\begin{figure*}
  \centering
  \includegraphics[width=\textwidth]{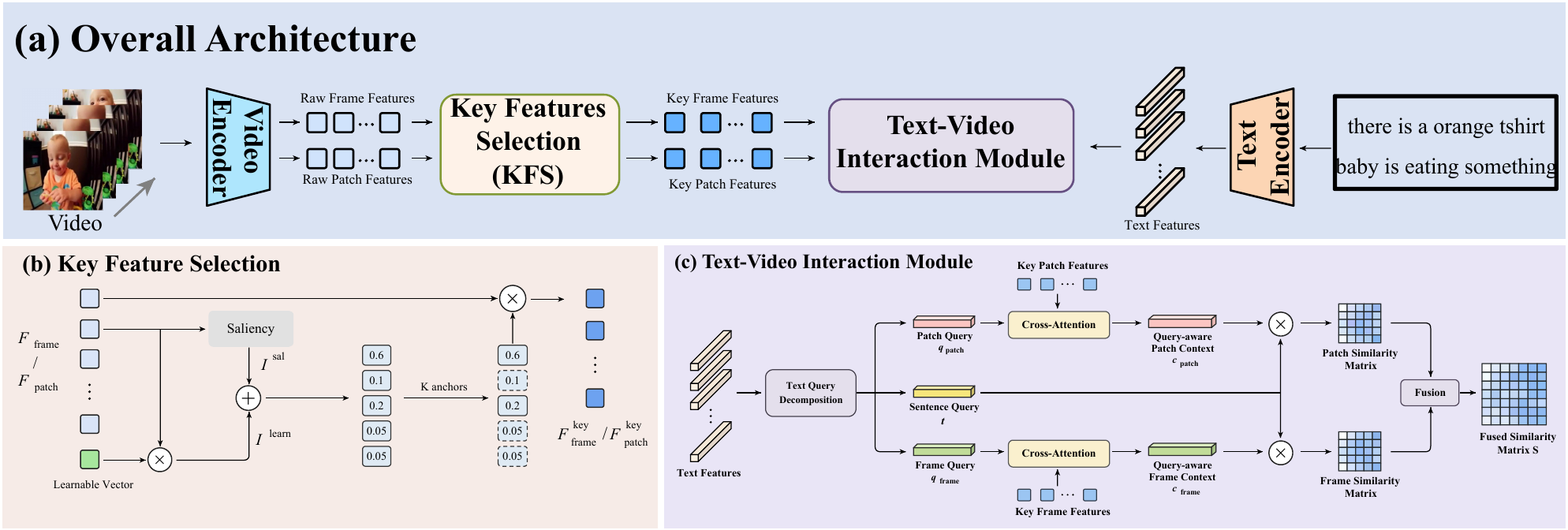}
  \caption{(a) Overview of the proposed MMTI. It consists of two core components: (b) Key Feature Selection (KFS) that mitigates visual redundancy by adaptively selecting informative frames and patches based on multi-dimensional saliency and learnable importance scores; (c) Text-Video Interaction Module~(TVIM) that decomposes the text query into sentence, frame, and patch queries (SFP) for multi-grained alignment with the selected key visual features.}
  \label{fig:methodOverview}
\end{figure*}

\subsection{Enhance Video Features from Text Interaction}
Based on CLIP, existing methods focus primarily on enhancing video representations through text interaction for retrieval~\cite{gorti2022x,li2023progressive,pei2023clipping,zhao2023cali}. For example, CLIP4Clip~\cite{luo2022clip4clip} and TS2-Net~\cite{liu2022ts2} use two independent encoders for vision and text. While efficient, the design often produces video representations that are overly abstract and lack multi-grained alignment with text queries. To address this gap, a line of research focuses on enhancing video representations through direct text interaction. For instance, X-CLIP~\cite{ma2022x} designs heavy cross-modal interaction blocks with multi-grained attention to learn joint representations. DGL~\cite{yang2024dgl} improves the performance of text-video retrieval by sharing a latent space to generate cross-modal prompts and introducing a global-local video attention mechanism with only a few parameter adjustments. To enrich visual and textual representations, recent methods have explored several strategies. Bima~\cite{le2025bima} fuses scene element features with the original visual features for visual debiasing, while NarVid~\cite{hur2025narrating} utilizes frame-level narration to enhance video features through cross-modal interaction, aiming to capture local details. However, these methods fail to exploit the inherent multi-level semantics in textual captions, thus limiting multi-grained cross-modal alignment~\cite{wu2023cap4video,jin2023diffusionret}.

\section{Method}
In this section, we present a detailed description of the proposed MMTI. We first define the problem, followed by an overview of the proposed method. Then, we elaborate on the key feature selection~(KFS) mechanism and text-video interaction module~(TVIM), where the text query is decomposed into SFP to achieve multi-grained text interaction. Finally, we present the objective function.

\subsection{Problem Definition}
The goal of text-video retrieval is to learn a joint embedding space where text queries and video content can be directly compared. Given a text query $q$ and a candidate video set 
\(
  \mathcal{V}=\{v_1,v_2,...,v_N\}
\), our objective is to find the video  \( v^{*} \) that is most semantically relevant to the query: 

\begin{equation}
  v^*=\arg\max_{v\in\mathcal{V}}s(q,v),
\end{equation}
where \( s(\cdot, \cdot) \) is a function that measures the text-video similarity. A video can be represented as a sequence of frames
\( \{f_1,f_2,...,f_T\} \), and each frame is further divided into P patches \( \{p_1,p_2,...,p_P\} \). The text query q is tokenized into a word sequence \( \{w_1,w_2,...,w_L\} \).

\subsection{Overview}
As shown in Figure~\ref{fig:methodOverview}, the overall architecture of MMTI consists of two core components. First, the key feature selection~(KFS) aims to effectively mitigate redundant information in videos. By evaluating both multi-dimensional saliency and learnable importance scores, KFS retains the key visual features, thereby compacting the original dense video features into a more discriminative representation while preserving key semantic information.

Second, in the text-video interaction module~(TVIM), the text query is decomposed into SFP. These decomposed queries then interact with the key video frame features and patch features, respectively, to generate text-relevant, enhanced visual context representations. This design enables MMTI to capture correspondences between text and visual content at different granularities.

Finally, during the training phase, we employ a joint objective function combining a bidirectional contrastive loss with a consistency regularization term. The contrastive loss aligns text and video representations through symmetric InfoNCE, while the consistency loss, implemented as KL divergence, encourages the fused cross-modal similarity to remain consistent with the ensemble of frame-level and patch-level similarities. During inference, the fused similarity scores are directly used for efficient text-video retrieval.

\begin{figure}[h]
  \centering
  \includegraphics[width=\linewidth]{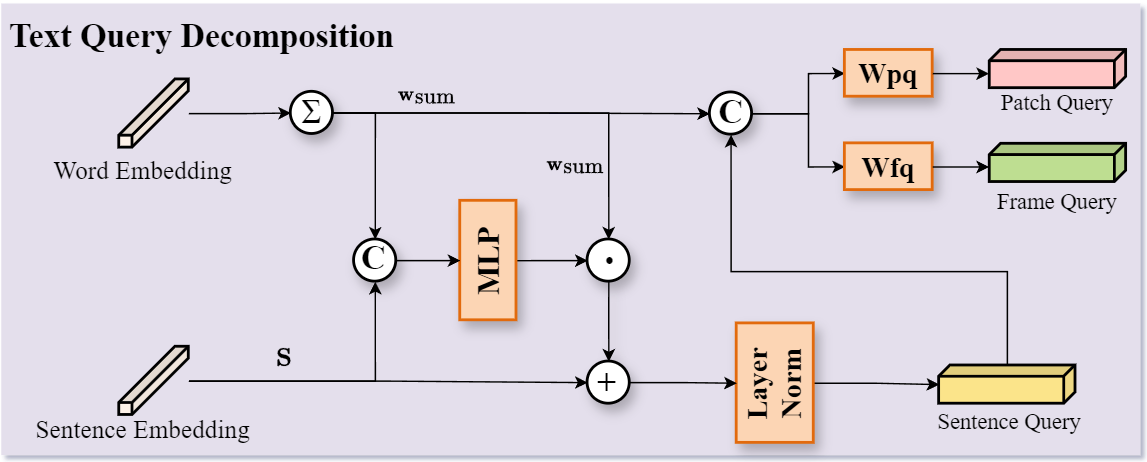}
  \caption{Pipeline of text query decomposition in TVIM. The text query is decomposed into three granularities: sentence
  query, frame query and patch query.}
  \label{fig:SFP}
\end{figure}

\subsection{Key Feature Selection}
Videos generally contain a large amount of information redundancy. Adjacent frames have similar content (inter-frame redundancy), and within a single frame, there are also many background or repetitive regions (intra-frame redundancy). Directly aligning these redundant visual features can lead to poor retrieval accuracy. To address this, we design a learnable key feature selection~(KFS) mechanism that aims to effectively compact frame-level and patch-level features while preserving key information.

\textbf{Frame-level Selection:} For an input video, we first extract frame-level features \(F_{\mathrm{frame}}\in\mathbb{R}^{B\times T\times D}\)
using the CLIP visual encoder, where $B$ is the batch size, $T$ is the number of frames, and $D$ is the feature dimension. The goal is to select and fuse the most representative frames. We combine multi-dimensional saliency and learnable importance scores to evaluate the significance of each frame. The multi-dimensional saliency \(I_t^{\mathrm{sal}}\) for frame $t$ is derived from its novelty, semantic distinctiveness, and energy variance:
\begin{equation}
      I_t^\mathrm{sal}=(1-\cos(\mathbf{f}_t,\mathbf{f}_{t-1}))+(1-\cos(\mathbf{f}_t,\bar{\mathbf{f}}))+\|\mathbf{f}_t-\bar{\mathbf{f}}_t\|_2,
\end{equation}
where \(\mathbf{f}_t\) is the feature of frame $t$, \(\bar{\mathbf{f}}\) is the global video representation, \(\bar{\mathbf{f}}_t\) is the mean of the feature vector. The learnable importance score is obtained via a projection:
\begin{equation}
    I_t^\mathrm{learn}=\mathbf{w}_\mathrm{frame}^\top\mathbf{f}_t,
\end{equation}
where \(  \mathbf{w}_{\mathrm{frame}} \) is a learnable weight vector.

The final importance score for each frame is an additive combination of the saliency and learned scores:
\begin{equation}
    I_t=\hat{I}_t^\mathrm{sal}+\hat{I}_t^\mathrm{learn},
\end{equation}
where \( \hat{I}_t^\mathrm{sal} \) and \(\hat{I}_t^{\mathrm{learn}}\) are standardized scores. Based on the resulting importance scores, we select the K highest-scoring frames as anchor frames, where \(K=\lceil T\cdot r_{\mathrm{frame}}\rceil\) and 
\( r_\mathrm{frame} \) is the key frame selection ratio. We fuse the selected frames with the original frames through a soft assignment mechanism. Let \( f_j^{\mathrm{anc}} \) denote the feature of the $j$-th anchor frame and \( I_j^{\mathrm{anc}} \) denote its corresponding combined importance score. We transform the anchor importance scores into a normalized anchor-prior distribution: 
\begin{equation}
\rho_j=\frac{\exp\left(I_j^\mathrm{anc}/\tau_\mathrm{anc}\right)}{\sum_{k=1}^K\exp\left(I_k^\mathrm{anc}/\tau_\mathrm{anc}\right)},
\end{equation}
where \( \tau_{\mathrm{anc}} \) controls the sharpness of the anchor-prior distribution. A larger \(\rho_{j}\) indicates that the corresponding anchor is considered more informative by the joint saliency and learnable importance assessment. We then compute a soft assignment matrix \(A\in\mathbb{R}^{T\times K}\) by jointly considering feature affinity and anchor importance:
\begin{equation}
    A_{i,j}=\frac{\rho_j\exp\left(\cos\left(f_i,f_j^\mathrm{anc}\right)/\tau_\mathrm{asg}\right)}{\sum_{k=1}^K\rho_k\exp\left(\cos\left(f_i,f_k^\mathrm{anc}\right)/\tau_\mathrm{asg}\right)},
\end{equation}
where \( \tau_{\mathrm{asg}} \) is the assignment temperature. The key frame features \(F_{\mathrm{frame}}^{\mathrm{key}}\in\mathbb{R}^{B\times K\times D}\)
are then obtained by aggregating all original frame features:
\begin{equation}
    F_{\mathrm{frame}}^{\mathrm{key}}=\left(\sum_{i=1}^TA_{i,j}\mathbf{f}_i\right)\oslash\left(\sum_{i=1}^TA_{i,j}\right),
\end{equation}
where \(\oslash\) denotes element-wise division, and the operations are applied for each batch and each selected frame index $j$.

\textbf{Patch-level Selection:} For each key frame, we similarly process its internal patch features \(F_{\mathrm{patch}}\in\mathbb{R}^{B\times T^{\prime}\times P\times D}\). We combine multi-dimensional saliency and learnable scores to select the most important patches. The saliency for a patch $p$ in a given frame is the following:
\begin{equation}
      I_p^\mathrm{sal}=(1-\cos(\mathbf{p},\mathbf{c}))+\|\mathbf{p}\|_2+(1-\cos(\mathbf{p},\mathbf{p}_\mathrm{left})),
\end{equation}
where \textbf{c} is the feature of the center patch, \( \mathbf{p}_{\mathrm{left}} \) is the feature of its left neighbor. The learnable importance score is \(I_p^\mathrm{learn}=\mathbf{w}_\mathrm{patch}^\top\mathbf{p}\). The final importance score is again a combination. The \(\mathrm{K'}\) patches with the highest scores are selected as anchor patches, where \(K^{\prime}=\lceil P\cdot r_{\mathrm{patch}}\rceil\) and \( r_{\mathrm{patch}}\) is the patch selection ratio. The key patch features are then generated using a similar soft assignment fusion mechanism as in frame selection, producing \( F_{\mathrm{patch}}^{\mathrm{key}}\in\mathbb{R}^{B\times T^{\prime}\times K^{\prime}\times D} \).

\subsection{Text-Video Interaction Module}
After the key visual features are obtained, the core component of MMTI—the text-video interaction module~(TVIM)—is responsible for performing multi-grained text-video alignment. The design of TVIM is motivated by the observation that existing methods overlook the granularity inherent in text features. MMTI decomposes the text query into queries at different granularities to more finely match the spatiotemporal structure of videos.

\textbf{Text Query Decomposition:} For an input text, we obtain a global sentence embedding \( \mathbf{s}\in\mathbb{R}^{D} \)
and word-level features \( \mathbf{W}\in\mathbb{R}^{L\times D} \)
using the CLIP text encoder. To enhance the sentence representation with fine-grained lexical information, we first compute a global summary:
\begin{equation}
    \mathbf{w}_{\mathrm{sum}}=\frac{1}{L-1}\sum_{i=2}^L\mathbf{w}_i,
\end{equation}
where \(\mathbf{w}_{i}\) denotes the feature of the $i$-th word token. We then fuse \textbf{s} and \(\mathbf{w}_{\mathrm{sum}}\) via a dynamic gating mechanism to produce the final sentence query, denoted by \textbf{t}, as follows:
\begin{equation}
    \mathbf{g}=\sigma\left(\mathbf{W}_g[\mathbf{s};\mathbf{w}_\mathrm{sum}]\right),
\end{equation}
\begin{equation}
    \mathbf{t}=\mathrm{LayerNorm}\left(\mathbf{s}+\mathbf{g}\odot\mathbf{w}_{\mathrm{sum}}\right),
\end{equation}
where \(\mathbf{W}_{g}\) is a linear projection. This refined text embedding \textbf{t} serves as the sentence query for computing global video-level similarity and is also used to generate the granular queries for frame-level and patch-level interactions. Subsequently, we construct two context-aware queries: \(\mathbf{q}_{\mathrm{frame}}=\mathrm{Normalize}\left(\mathbf{W}_{\mathrm{fq}}[\mathbf{t};\mathbf{w}_{\mathrm{sum}}]\right)\), \(\mathbf{q}_{\mathrm{patch}}=\mathrm{Normalize}\left(\mathbf{W}_{\mathrm{pq}}[\mathbf{w}_{\mathrm{sum}};\mathbf{t}]\right)\),
where \(\mathbf{W}_{\mathrm{fq}}\) and \(\mathbf{W}_{\mathrm{pq}}\) are linear projections, and Normalize denotes L2 normalization. As shown in Figure~\ref{fig:SFP}, the text embedding is decomposed into three different granularities.

\textbf{Multi-grained Cross-modal Interaction:} We apply a shared confidence-modulated attention mechanism at both frame and patch granularities, with the query, key features, and projection weights adapted to each level.

At the \textbf{frame level}, the frame query \(\mathbf{q}_{\mathrm{frame}}\) interacts with the key frame features \(F_{\mathrm{frame}}^{\mathrm{key}}\in\mathbb{R}^{B\times T^{\prime}\times D}\). First, we compute a confidence score for each frame to assess its relevance to the text:
\begin{equation}
    c_{\mathrm{frame}}^j=\sigma(\mathbf{W}_{\mathrm{conf}}\mathbf{f}_j^{\mathrm{key}}),
\end{equation}
where \(\mathbf{W}_{\mathrm{conf}}\) is a learnable projection. The attention weight for each frame is then computed by modulating the dot-product similarity with the confidence score:
\begin{equation}
    \alpha_j=\frac{\exp((\mathbf{q}_\mathrm{frame}\cdot\mathbf{f}_j^\mathrm{key})/\tau+\log c_\mathrm{frame}^j)}{\sum_{k=1}^{T^{\prime}}\exp((\mathbf{q}_\mathrm{frame}\cdot\mathbf{f}_k^\mathrm{key})/\tau+\log c_\mathrm{frame}^k)},
\end{equation}
where \(\tau\) is the temperature. The query-aware frame context feature is obtained via weighted summation:
\begin{equation}
    \mathbf{c}_{\mathrm{frame}}=\sum_{j=1}^{T^{\prime}}\alpha_{j}\mathbf{f}_{j}^{\mathrm{key}},
\end{equation}
Finally, the frame-level similarity score can be computed as:
\begin{equation}
    S_\mathrm{frame}=\gamma\cdot(\mathbf{t}\cdot\mathbf{c}_\mathrm{frame}),
\end{equation}
where \(\gamma\) is the learned logit scale from the CLIP model.

The \textbf{patch-level} interaction follows an analogous procedure. The patch query \(\mathbf{q}_{\mathrm{patch}}\) interacts with the key patch features \(F_{\mathrm{patch}}^{\mathrm{key}}\in\mathbb{R}^{B\times T^{\prime}\times P^{\prime}\times D}\) through the same confidence-modulated attention formulation. For efficient computation, a chunk-wise computation strategy is employed to handle the larger number of patches.

\subsection{Objective Function}

The overall loss combines a bidirectional InfoNCE loss and an auxiliary consistency loss:

\begin{equation}
\mathcal{L} = \mathcal{L}_n + \lambda \mathcal{L}_c,
\end{equation}
where $\lambda$ is the balancing weight. The InfoNCE loss is computed bidirectionally:

\begin{equation}
\mathcal{L}_{t2v} = -\frac{1}{B}\sum_{i=1}^{B}\log\frac{e^{S_{i,i}/\tau}}{\sum_{j}e^{S_{i,j}/\tau}},
\end{equation}

\begin{equation}
\mathcal{L}_{v2t} = -\frac{1}{B}\sum_{i=1}^{B}\log\frac{e^{S_{i,i}/\tau}}{\sum_{j}e^{S_{j,i}/\tau}},
\end{equation}

\begin{equation}
\mathcal{L}_n = \frac{1}{2}\big(\mathcal{L}_{t2v} + \mathcal{L}_{v2t}\big),
\end{equation}
where $S$ is the fused similarity matrix from TVIM. The consistency loss regularizes the fused similarity towards the ensemble of the two similarities at different granularities:

\begin{equation}
\mathcal{L}_c^{\mathrm{row}} = \frac{1}{B}\sum_{i=1}^{B}\mathrm{KL}\big(P_t^{(i)}\|P_s^{(i)}\big),
\end{equation}

\begin{equation}
\mathcal{L}_c^{\mathrm{col}} = \frac{1}{B}\sum_{i=1}^{B}\mathrm{KL}\big(P_t^{(i)\top}\|P_s^{(i)\top}\big),
\end{equation}

\begin{equation}
\mathcal{L}_c = \frac{1}{2}\big(\mathcal{L}_c^{\mathrm{row}} + \mathcal{L}_c^{\mathrm{col}}\big),
\end{equation}
where $P_t = \mathrm{softmax}((S_{\mathrm{frame}}+S_{\mathrm{patch}})/(2\tau_c))$, $P_s = \mathrm{softmax}(S/\tau_c)$, and $\tau_c$ is the temperature.

% ------------------------------------------------------------
%  table 1: MSR-VTT + ActivityNet + DiDeMo
% ------------------------------------------------------------
\begin{table*}[h]
  \centering
  \small
  \setlength{\tabcolsep}{2.8pt}
  \begin{tabular}{l@{\hskip 4pt}|cccc|cccc|cccc}
    \toprule
    \multirow{2}{*}{Methods} & \multicolumn{4}{c|}{MSR-VTT} & \multicolumn{4}{c|}{ActivityNet} & \multicolumn{4}{c}{DiDeMo} \\
    \cmidrule(lr){2-5} \cmidrule(lr){6-9} \cmidrule(lr){10-13}
    & R@1$\uparrow$ & R@5$\uparrow$ & R@10$\uparrow$ & Rsum$\uparrow$ & R@1$\uparrow$ & R@5$\uparrow$ & R@10$\uparrow$ & Rsum$\uparrow$ & R@1$\uparrow$ & R@5$\uparrow$ & R@10$\uparrow$ & Rsum$\uparrow$ \\
    \midrule
    HBI~\cite{jin2023video}             & 48.6 & 74.6 & 83.4 & 206.6 & 42.2 & 73.0 & 84.6 & 199.8 & 46.9 & 74.9 & 82.7 & 204.5 \\
    DiCoSA~\cite{jin2023text}           & 47.5 & 74.7 & 83.8 & 206.0 & 42.1 & 73.6 & 84.6 & 200.3 & 45.7 & 74.6 & 83.5 & 203.8 \\
    UATVR~\cite{fang2023uatvr}          & 47.5 & 73.9 & 83.5 & 204.9 & --   & --   & --   & -- & 43.1 & 71.8 & 82.3 & 197.2 \\
    DiffusionRet~\cite{jin2023diffusionret} & 49.0 & 75.2 & 82.7 & 206.9 & 45.8 & 75.6 & 86.3 & 207.7 & 46.7 & 74.7 & 82.7 & 204.1 \\
    Cap4Video~\cite{wu2023cap4video}    & 49.3 & 74.3 & 83.8 & 207.4 & --   & --   & --   & -- & --   & --   & --   & -- \\
    DGL~\cite{yang2024dgl}              & 45.8 & 69.3 & 79.4 & 194.5 & 38.6 & 69.2 & 81.6 & 189.4 & --   & --   & --   & -- \\
    EERCF~\cite{tian2024towards}        & 47.8 & 74.1 & 84.1 & 206.0 & 43.1 & 74.5 & 86.0 & 203.6 & --   & --   & --   & -- \\
    TeachCLIP~\cite{tian2024holistic}   & 46.8 & 74.3 & 82.6 & 203.7 & 42.2 & 72.7 & 85.2 & 200.1 & 43.7 & 71.2 & 81.1 & 196.0 \\
    DITS~\cite{wang2024diffusion}       & 51.9 & 75.7 & 84.6 & 212.2 & --   & --   & --   & -- & 51.1 & 77.9 & 85.8 & 214.8 \\
    MPT~\cite{zhang2024mpt}             & 46.3 & 70.9 & 80.7 & 197.9 & 41.4   & 70.9   & 82.9   & 195.2 & 46.4   & 72.2   & 81.4   & 200.0  \\
    ProTA~\cite{fang2024prota}          & 48.1 & 75.4 & 84.3 & 207.8 & --   & --   & --   & -- & 47.2   & 74.6   & 83.0   & 204.8 \\
    T-MASS~\cite{wang2024text}          & 50.2 & 75.3 & 85.1 & 210.6 & --   & --   & --   & -- & 50.9   & 77.2  & 85.3  & 213.4 \\
    TempMe~\cite{shen2024tempme}        & 46.1 & 71.8 & 80.7 & 198.6 & 44.9 & 75.2 & 85.5 & 205.6 & 48.0 & 72.4 & 81.8 & 202.2 \\
    TextProxy~\cite{xiao2025text}       & 52.3 & 77.8 & 85.8 & 215.9 & 53.0 & 80.9 & 89.6 & 223.5 & 50.6 & 76.9 & 86.0 & 213.5 \\
    NarVid~\cite{hur2025narrating}      & 51.0 & 76.4 & 85.2 & 212.6 & --   & --   & --   & -- & 53.4 & 79.1 & 86.3 & 218.8 \\
    VTC~\cite{liu2025queries}           & 53.1 & 78.5 & \underline{88.5} & \underline{220.1} & -- & -- & -- & -- & 52.2 & 79.3 & 87.4 & 218.9 \\
    V-Sparse~\cite{liu2026v}       	& 50.9 & 76.2 & 85.3 & 212.4 & 45.1 & 74.5 &  85.0  & 204.6 & 51.2 & 78.1 & 86.1 & 215.4 \\
    Bima~\cite{le2025bima}              & 53.5 & 78.6 & 86.5 & 218.6 & 55.4 & \underline{83.4} & \textbf{92.4} & \underline{231.2} & \underline{56.0} & \underline{81.9} & \underline{87.8} & \underline{225.7} \\
    AMD-Net~\cite{meng2026appearance}   & \underline{55.2} & \underline{78.7} & 85.3 & 219.2 & \textbf{57.8}   & 82.0   & 90.1   & 229.9 & 53.5 & 78.9 & 86.9 & 219.3 \\
    \midrule
    \textbf{MMTI (Ours)} & \textbf{56.6} & \textbf{89.4} & \textbf{95.0} & \textbf{241.0} & \underline{56.0} & \textbf{85.0} & \underline{92.3} & \textbf{233.3} & \textbf{59.5} & \textbf{87.5} & \textbf{93.9} & \textbf{240.9} \\
    \bottomrule
    \end{tabular}
  \caption{Text-to-video comparison on MSR-VTT, ActivityNet, and DiDeMo datasets. All results are reported under the ViT-B/32 backbone. The best and second best are \textbf{bold} and \underline{underlined}.}
  \label{tab:main_comparison}
\end{table*}

\section{Experiments}
\subsection{Experimental Settings}
\textbf{Datasets:} We conducted experiments on four major text-video retrieval
benchmarking datasets: MSR-VTT~\cite{xu2016msr} contains 10,000 videos, each accompanied by 20 text captions, and is evaluated using the commonly used 1k-A test division. MSVD~\cite{chen2011collecting} contains 1,970 videos, each with an average of 40 captions. We follow the official split with 1,200 videos for training and 670 videos for testing. ActivityNet Captions~\cite{krishna2017dense} consists of 20,000 YouTube videos. Following the training and evaluation protocol from~\cite{luo2022clip4clip}, we present results on the ``val'' split, using 10,009 videos as the training set and 4,917 as the test set, respectively. DiDeMo~\cite{anne2017localizing} contains 10,000 videos with 40,000 captions.

\textbf{Implementation Details:} MMTI is initialized with the pre-trained weights of CLIP (ViT-B/32). The maximum number of frames is set to 36. The maximum word is set to 32. We use the Adam~\cite{kingma2014adam} optimizer with an initial learning rate of 1e-4 and a cosine annealing schedule. The batch size is set to 256, and training is performed on 4 NVIDIA A100 GPUs. Unless otherwise specified, models are trained for 5 epochs; DiDeMo and ActivityNet are trained for 10 and 20 epochs, respectively. For reproducibility, all experiments are implemented using the PyTorch framework. We set the frame selection ratio to 0.67 and the patch selection ratio to 0.55. For fair comparison, all baseline results are reported under the ViT-B/32 backbone with the same dataset splits and evaluation protocols.

\textbf{Evaluation Metrics:} We adopt standard retrieval metrics: Recall at K (R@1, R@5, R@10) and Rsum (R@1 + R@5 + R@10) as primary metrics for quantitative assessment.

% ------------------------------------------------------------
% table : MSVD
% ------------------------------------------------------------
\begin{table}[t]
  \centering
  \small
  \setlength{\tabcolsep}{3pt}
  \begin{tabular}{l|cccc}
    \toprule
    \multirow{2}{*}{Methods} & \multicolumn{4}{c}{MSVD} \\
    \cmidrule(lr){2-5}
    & R@1$\uparrow$ & R@5$\uparrow$ & R@10$\uparrow$ & Rsum$\uparrow$ \\
    \midrule
    DiCoSA~\cite{jin2023text}         & 47.4 & 76.8 & 86.0 & 210.2 \\
    UATVR~\cite{fang2023uatvr}        & 46.0 & 76.3 & 85.1 & 207.4 \\
    DiffusionRet~\cite{jin2023diffusionret} & 46.6 & 75.9 & 84.1 & 206.6 \\
    EERCF~\cite{tian2024towards}      & 47.0 & 77.5 & 85.4 & 209.9 \\
    TeachCLIP~\cite{tian2024holistic} & 47.4 & 77.3 & 85.5 & 210.2 \\
    NarVid~\cite{hur2025narrating}    & 53.1 & 81.4 & 88.8 & 223.3 \\
    Bima~\cite{le2025bima}            & 55.7 & 83.2 & \underline{91.2} & 230.1 \\
    AMD-Net~\cite{meng2026appearance} & \textbf{56.8} & \underline{84.0} & 90.2 & \underline{231.0} \\
    \midrule
    \textbf{MMTI (Ours)} & \underline{56.1} & \textbf{85.4} & \textbf{93.1} & \textbf{234.6} \\
    \bottomrule
  \end{tabular}
  \caption{Text-to-video comparison on MSVD dataset. The best and second best are \textbf{bold} and \underline{underlined}.}
  \label{tab:msvd_comparison}
\end{table}

\subsection{Comparison with State-of-the-Art Methods}
Table~\ref{tab:main_comparison} and~\ref{tab:msvd_comparison} report the comparison on MSR-VTT, ActivityNet, DiDeMo, and MSVD. Our MMTI achieves state-of-the-art performance under the ViT-B/32 backbone. MMTI obtains 56.6\% in R@1, 89.4\% in R@5, and 95.0\% in R@10, with an Rsum of 241.0 on MSR-VTT, outperforming all previous methods. Moreover, compared to AMD-Net, MMTI achieves R@5 and R@10 improvements of over 10\% and over 9\%, respectively. These significant gains demonstrate the effectiveness of mitigating visual redundancy through KFS and achieving multi-grained alignment via SFP.

Notably, MMTI exhibits particularly pronounced gains in R@5 and R@10 across all benchmarks, while R@1 remains consistently competitive or best. We attribute this pattern to the text-agnostic nature of KFS: unlike text-conditioned compression strategies that may capture dataset-specific textual biases, our multi-dimensional saliency assessment operates purely within the visual modality, yielding compact and unbiased representations that generalize robustly. The substantial R@5 and R@10 improvements indicate that redundancy removal cleans the candidate ranking, allowing truly relevant videos to concentrate at higher ranks. On ActivityNet, MMTI trails AMD-Net slightly in R@1 yet still achieves the best Rsum (233.3), further confirming that vision-driven redundancy mitigation provides complementary gains.

\begin{table}[t]
\centering
\small
\setlength{\tabcolsep}{2pt}
\begin{tabular}{l c c c c c c c}
\toprule
Variant & SFP & KL & Agg. & KFS & R@1$\uparrow$ & R@5$\uparrow$ & R@10$\uparrow$ \\
\midrule
\#1 w/o SFP      &   & $\checkmark$ & Soft & N+S+E & 48.0 & 80.1 & 89.5 \\
\#2 w/o KFS      & $\checkmark$ & $\checkmark$ & ---  & ---   & 50.2 & 81.3 & 89.9 \\
\#3 w/o KL       & $\checkmark$ &   & Soft & N+S+E & 54.1 & 87.4 & 94.0 \\
\#4 w/o Novelty  & $\checkmark$ & $\checkmark$ & Soft & S+E   & 47.7 & 79.7 & 89.2 \\
\#5 w/o Semantic & $\checkmark$ & $\checkmark$ & Soft & N+E   & 47.7 & 80.7 & 89.6 \\
\#6 w/o Energy   & $\checkmark$ & $\checkmark$ & Soft & N+S   & 47.2 & 80.1 & 89.8 \\
\#7 Hard TopK    & $\checkmark$ & $\checkmark$ & TopK & N+S+E & 48.4 & 80.2 & 89.1 \\
\midrule
\rowcolor{gray!15}
\#8 Full         & $\checkmark$ & $\checkmark$ & Soft & N+S+E & \textbf{56.6} & \textbf{89.4} & \textbf{95.0} \\
\bottomrule
\end{tabular}
\caption{Ablation study of MMTI components on MSR-VTT under the text-to-video setting. ``KL'' denotes the consistency loss. ``Agg.'' denotes the aggregation method in KFS. ``KFS'' shows which saliency components are enabled: N (Novelty), S (Semantic Distinctiveness), E (Energy Variance).}
\label{tab:ablation_extended}
\end{table}

\subsection{Ablation Study}
To investigate the contribution of each component in MMTI, we conduct ablation experiments on MSR-VTT. Table~\ref{tab:ablation_extended} summarizes the results under the text-to-video setting.

\textbf{Effect of SFP.} Removing the SFP and replacing it with a single global text embedding reduces R@1 from 56.6\% to 48.0\% and Rsum from 241.0 to 217.6. This indeed confirms that decomposing text queries into three granularities is essential for aligning with the hierarchical structure of video frames and patches, allowing the model to capture both global semantics and local details.

\textbf{Effect of KFS.} When KFS is completely removed, all frames and patches are retained, leading to a performance drop to 50.2\% R@1 and 221.4 Rsum. This indicates that visual redundancy not only harms efficiency but also degrades retrieval accuracy by introducing repetitive information. To further analyze KFS, we ablate each saliency component individually. Removing the novelty (w/o Novelty) reduces R@1 to 47.7\%, while removing semantic distinctiveness (w/o Semantic) yields 47.7\% R@1, and removing energy variance (w/o Energy) drops R@1 to 47.2\%. The substantial degradation across all three variants demonstrates that each saliency component captures complementary aspects of visual importance and jointly contributes to identifying discriminative visual content. Furthermore, replacing the soft assignment fusion with hard TopK selection (Hard TopK) achieves only 48.4\% R@1, far lower than 56.6\% of the full model, which highlights the importance of the differentiable fusion mechanism for effective feature aggregation.

\textbf{Effect of loss functions.} Removing the consistency loss (w/o KL) causes R@1 to drop to 54.1\% and Rsum to 235.5, demonstrating that the KL divergence regularization helps maintain consistent cross-modal alignment across different granularities. The full MMTI model achieves the best performance across all metrics, which further validates the complementary benefits of all proposed components.

\begin{figure}[t]
  \centering
  \includegraphics[width=\linewidth]{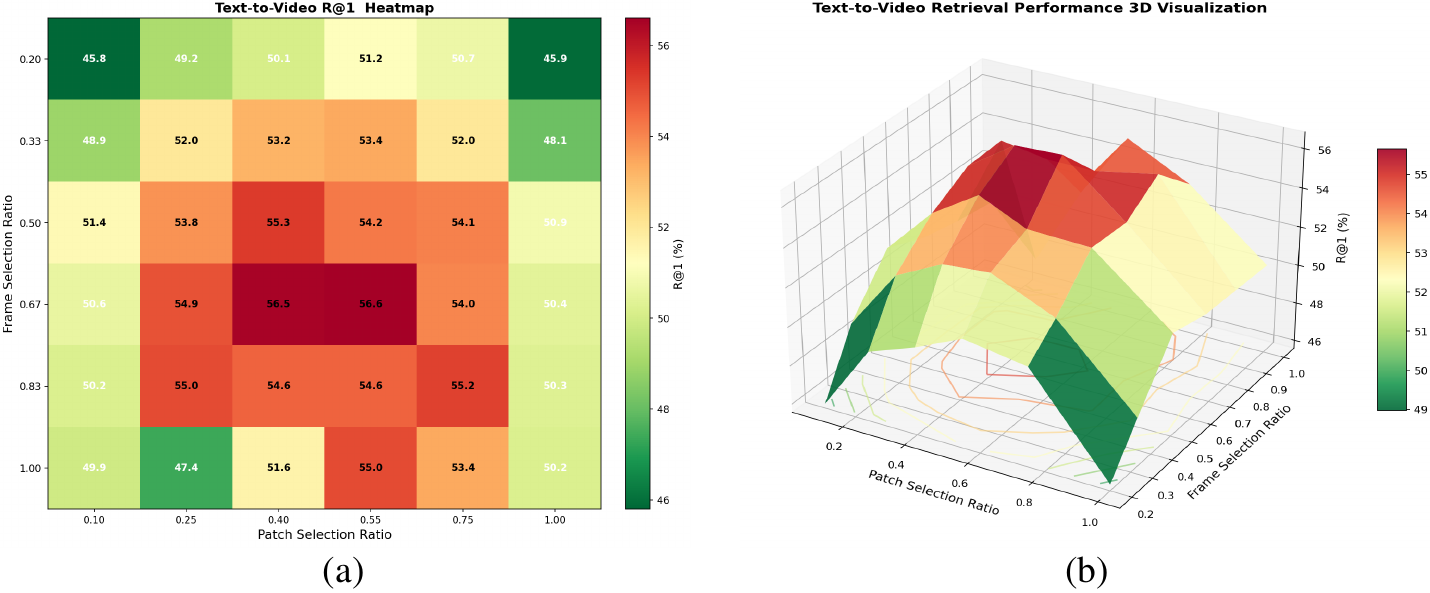}
  \caption{Selection ratio analysis of KFS on MSR-VTT. (a) Heatmap of R@1 under different frame and patch selection ratios. (b) 3D surface visualization of the joint effect of frame and patch selection ratios on retrieval performance.}
  \label{fig:selection_ratio_analysis}
\end{figure}

\subsection{KFS Selection Ratio Analysis}
To investigate the impact of the key feature selection mechanism, we systematically vary the frame selection ratio \(r_{\mathrm{frame}}\) and patch selection ratio \(r_{\mathrm{patch}}\) on MSR-VTT. Figure~\ref{fig:selection_ratio_analysis} comprehensively visualizes the results: (a) a heatmap of R@1 under different ratio combinations, (b) a 3D surface visualization of the joint effects.

The optimal configuration is \(r_{\mathrm{frame}}=0.67, r_{\mathrm{patch}}=0.55\), achieving 56.6\% R@1. Compared to the baseline without KFS, this yields a +6.4\% improvement, validating that visual redundancy degrades retrieval accuracy and KFS effectively mitigates it by compacting dense visual features. Moderate selection consistently benefits performance: configurations with \(r_{\mathrm{frame}}\in[0.50,0.83]\) and \(r_{\mathrm{patch}}\in[0.25,0.75]\) all outperform the baseline, indicating that retaining a reasonable portion of informative visual content while discarding redundant information is beneficial. However, overly aggressive selection (e.g., \(r_{\mathrm{frame}}=0.20, r_{\mathrm{patch}}=0.10\)) drops R@1 to 45.8\%, suggesting that excessive compression loses discriminative semantics essential for alignment. The results further show that both selection ratios jointly influence the final performance, and the model achieves consistent gains across a reasonably broad operating range. The optimal configuration transfers well across datasets and is adopted uniformly across all four benchmarks without per-dataset tuning.

\subsection{Qualitative Analysis}
\begin{figure}[t]
  \centering
  \includegraphics[width=\linewidth]{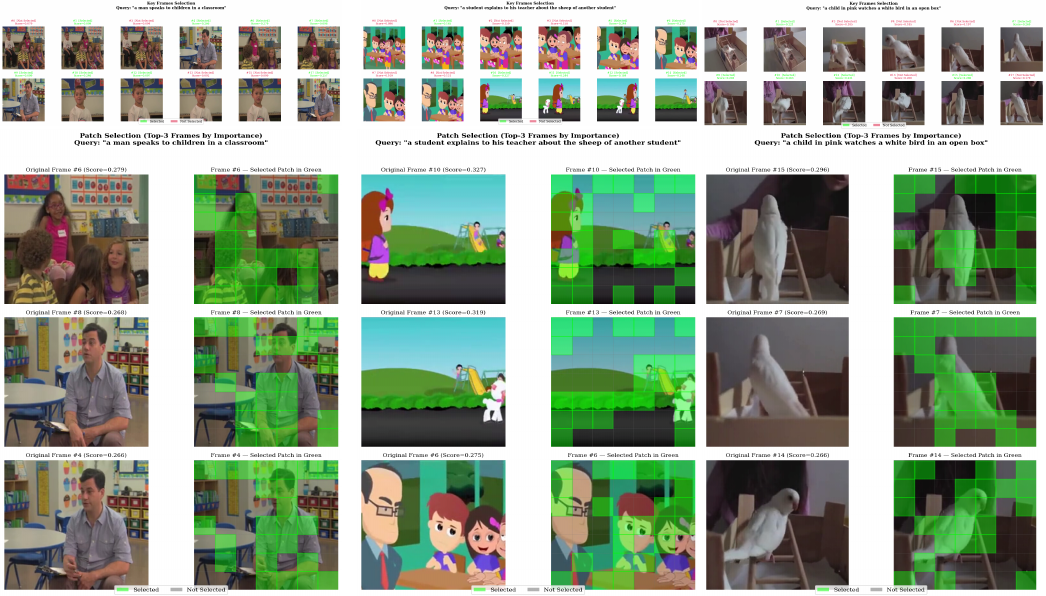}
  \caption{KFS selection visualization. Selected frames and patches are marked in green.}
  \label{fig:KFS_vis}
\end{figure}

To further demonstrate the effectiveness of KFS, we visualize the selected key frames and key patches in Figure~\ref{fig:KFS_vis} on three representative text queries, with selected regions marked in green. For the query ``a man speaks to children in a classroom'', KFS accurately identifies frames containing the man and children as key subjects and selects patches covering their interactive regions, while suppressing irrelevant background patches. For ``a student explains to his teacher about the sheep of another student'', KFS identifies frames where all four key entities---the student, teacher, another student, and sheep---are present and selects patches covering each entity, demonstrating the ability to handle complex multi-object scenes. For ``a child in pink watches a white bird in an open box'', KFS selects frames depicting the child and the box while retaining patches of the pink-clad child, the white bird, and the box, showing fine-grained discrimination of attribute-specific objects. Across all three queries, KFS consistently selects informative frames and patches that correspond to the key semantics described in the text, demonstrating that the joint evaluation of multi-dimensional saliency and learnable importance scores effectively identifies discriminative visual content for accurate cross-modal alignment.

\section{Conclusions}
In this paper, we propose MMTI, a text-video retrieval method that jointly mitigates visual redundancy and enables multi-grained text-video interaction. A learnable key feature selection~(KFS) mechanism adaptively identifies and aggregates informative frames and patches by jointly evaluating multi-dimensional saliency and learnable importance scores, effectively compacting dense visual features and mitigating visual redundancy. Furthermore, we introduce a text-video interaction module~(TVIM) that decomposes the text query into sentence, frame, and patch queries~(SFP) for multi-grained alignment with the selected key visual features, capturing correspondences at complementary semantic levels. Extensive experiments on four standard benchmarks demonstrate that our method outperforms the state-of-the-art.

\bibliography{aaai2027}

% Check whether the conference requires a reproducibility checklist to be included in the paper.
% If so, you can uncomment the following line and ajust the path to include it.
% \input{ReproducibilityChecklist.tex}

\end{document}